%% file: article_4.tex
\pdfoutput=1
\documentclass[
10pt, 
a4paper, 
oneside, 
headinclude,footinclude, 
BCOR5mm, 
]{scrartcl}

\input{structure.tex} 

\usepackage[ddmmyyyy]{datetime}

\title{\normalfont\spacedallcaps{Protocolo de Revisão Sistemática da Literatura para Wireless Mesh Networks com mecanismos de incentivo financeiro}} 

\author{\spacedlowsmallcaps{Rodolfo B. S. Carvalho}} 

\date{\today} 

\begin{document}


\renewcommand{\sectionmark}[1]{\markright{\spacedlowsmallcaps{#1}}} 
\lehead{\mbox{\llap{\small\thepage\kern1em\color{halfgray} \vline}\color{halfgray}\hspace{0.5em}\rightmark\hfil}} 

\pagestyle{scrheadings} 


\maketitle 






\section*{Objetivo desejado com este estudo:} 

( ) Capítulo de estado da arte da dissertação/tese
\\( ) Publicar um paper com o resultado em uma conferência
\\( ) Publicar um paper com o resultado em um journal
\\(x) Publicar o resultado como parte de um paper em uma conferência ou journal







\section{Justificativa da Necessidade}
As Wireless Mesh Networks (WMNs) são redes capazes não apenas de fornecer conectividade sem fio a largas áreas (como shoppings e centros urbanos), mas também levar tal conectividade a locais muito mais remotos do território \cite{AKYILDIZ2005445}, principalmente pela redução de preços em aparelhos e facilidade em sua implantação, levando o surgimento de diversas redes comunitárias \cite{BAIG2015150}, que ajudam a resolver o problema de última milha, onde áreas mais remotas não possuem acesso a rede de internet por falta de interesse econômico das grandes operadoras de telefonia.

Os altos custos para implantação da rede cabeada, ainda no ano de 2017, justificavam a carência de acesso estável à rede de internet por mais da metade da população mundial \cite{brown20172017}. Tais problemas de custos, apesar de serem contornados pela utilização das WMNs, estas ainda sofrem com a sustentação de seus serviços, buscando meios de incentivar seu uso, manutenção e possíveis expansões, sem que dependa de iniciativas altruístas. 

Afim de garantir a correta sustentabilidade das WMNs, surgiram nos últimos anos alguns trabalhos, dentro da literatura, que abordam a utilização de mecanismos para incentivo financeiro atrelados à rede \cite{8311658}. Como maneira de julgar o esforço e qualidade do trabalho desenvolvido pela rede, os mecanismos de incentivo buscam sempre estar associados aos protocolos de roteamento, presentes na camada de rede.

Existem diversos trabalhos que citam protocolos de roteamento associados a características especificamente voltadas para WMNs \cite{7573902}\cite{8746361}, buscando otimizar o funcionamento das mesmas e contornando problemas específicos, entretanto, a aplicação de mecanismos de incentivo financeiro aos protocolos ainda não possui devida sumarização dos estudos existentes, gerando para isso a necessidade de uma análise mais profunda da literatura. 

Alguns trabalhos como \cite{8311658} e \cite{10_1145_3211933_3211936} citam a utilização de mecanismos de incentivo por meio das Blockchains associadas às WMNs, mas não se aprofundam quanto ao funcionamento dos protocolos para otimização do roteamento na rede, sendo necessário um estudo a respeito dos trabalhos na área, buscando entender o atual patamar de avanço das pesquisas, apoiando o correto caminho para desenvolvimento dos futuros esforços, e colaborando para o aprimoramento da sustentabilidade das WMNs.
%
%
 

\section{Objetivo e questão(ões) de pesquisa}
O objetivo desta revisão sistemática é avaliar o estado da arte sobre protocolos de roteamento voltados para WMNs, quando associados a mecanismos para incentivo financeiro que visam garantir a auto-sustentabilidade da rede, buscando identificar as melhorias e riscos associados assim como a eficiência dos mecanismos adotados dentro das redes, garantindo assim o aumento da confiabilidade na utilização de tais protocolos para roteamento de WMNs incentivadas.

Perguntas de pesquisa:

\begin{itemize}
	\item Como a utilização dos protocolos associados a mecanismos de incentivo financeiro tem sido adotados para WMNs? 
	\item Quais as técnicas mais utilizadas para implementação de mecanismos de incentivo para WMNs dentro de protocolos de roteamento? 
	\item Qual a eficiência dos modelos propostos em testes realizados para o incentivo de WMNs?
	\item Quais os benefícios e melhorias podem ser alcançadas para as WMNs com a utilização dos protocolos adaptados com mecanismos de incentivo financeiro?
	\item Quais os custos e riscos associados a utilização dos protocolos adaptados com mecanismos de incentivo financeiro para as WMNs?
\end{itemize}



%
%
%
%
%
%
%
%
%
%
%


\section{String de busca e fontes de pesquisa}
A fim de identificar os estudos que mais se relacionam ao assunto abordado por esta RSL, será utilizado como string de busca:

\textit{(protocol OR protocolo) AND (WMN OR "wireless mesh network" OR "redes de malha sem fio") AND (incentive OR economic OR money OR credit OR incentivo OR econômico OR dinheiro OR crédito)}

 O critério utilizado para a seleção de estudos baseou-se na pesquisa de trabalhos dentro de 5 bases de dados (IEEE, ACM, Wiley, Springer, Elsevier), por terem uma grande cobertura e serem extensamente utilizadas para produção científica na área de Ciência da Computação. Durante a análise preliminar dos estudos encontrados em cada base, foram identificadas as seguintes quantidades:
 \begin{itemize}
 	\item IEEE: 4 estudos.
 	\item ACM: 58 estudos.
 	\item Wiley: 47 estudos.
 	\item Springer: 171 estudos.
 	\item Elsevier: 146 estudos.
 \end{itemize}
Totalizando 426 estudos incluindo possíveis duplicações (busca atualizada na data de 29/10/2020).
 
 Foram incluídos também como fonte de pesquisa trabalhos identificados através do procedimento metodológico \textit{snowball}(bola de neve), realizados sobre os estudos iniciais, identificados nas bases de dados.
 Também foram considerados cinco trabalhos sugeridos por profissionais envolvidos com o tema abordado, e que fazem parte do laboratório de pesquisa utilizado durante o estudo.

\section{Critérios de inclusão e exclusão}

\subsection*{Critérios de Inclusão}
\begin{itemize}
	\item Trabalhos que abordem a utilização de protocolos de roteamento dentro de WMNs.
	\item Trabalhos que adotem mecanismos de incentivo para o correto roteamento de pacotes dentro da rede.
	\item Trabalhos publicados desde o ano de 2017 até a atual data de realização deste trabalho.
	\item Trabalhos de artigos acadêmicos publicados em anais de eventos e revistas científicas.
	\item Trabalhos escritos em língua Inglesa e Portuguesa.
\end{itemize}
\subsection*{Critérios de Exclusão}
\begin{itemize}
	\item Serão excluídos trabalhos que adotem mecanismos de reputação como critério exclusivo para roteamento de pacotes.
	\item Serão excluídos trabalhos que abordem apenas ambientes de redes MANETs.
	\item Serão excluídos trabalhos que não abordem propostas de solução para mecanismos de incentivo em redes WMNs.
	\item Serão excluídos trabalhos publicados como artigos curtos ou pôsteres.
\end{itemize}

\section{Avaliação da qualidade}
A avaliação de qualidade dos estudos baseia-se na estrutura apresentada por \cite{kitchenham2007guidelines}, onde é realizada uma pontuação que utiliza critérios estabelecidos na forma de perguntas, para cada pergunta a resposta "sim" equivale a 1 ponto, e respostas não claras mas tendendo a sim recebem 0.5 ponto. A soma de todos os itens marcados afirmativamente constituem a pontuação total, onde pontuações no intervalo de 0-2.5 serão considerados trabalhos de qualidade pobre, de 3-5.5 como qualidade baixa, de 6-8.5 como qualidade satisfatória e de 9-11 como qualidade alta. As perguntas adotadas para análise dos trabalhos foram:

\begin{itemize}
	\item Os objetivos das questões de pesquisa estão claramente definidos?
	\item As amostras tomadas para sustentação do estudo representam satisfatoriamente a generalidade presente no ambiente real?
	\item Existem algum grupo/categoria de comparação para os resultados apresentados?
	\item O trabalho apresenta uma descrição clara e adequada dos métodos de coleta de dados?
	\item O trabalho apresenta uma descrição clara e adequada dos métodos de análise de dados?
	\item As possíveis hipóteses estatísticas foram levadas em consideração?
	\item Todas as indagações colocadas pelo estudo foram respondidas?
	\item As conclusões encontradas estão claramente apresentadas e relacionadas aos objetivos de pesquisa do estudo?
	\item O estudo apresenta testes virtuais ou reais comparativos com outros grupos/categorias possíveis?
	\item Os parâmetros utilizados para configuração e realização dos testes foram claramente descritos?
	\item As conclusões do estudo permitem que as questões de pesquisa sejam respondidas?
\end{itemize}

\section{Procedimento de seleção}
O processo de seleção ocorrerá conforme descrito pelas seguintes fases:
\begin{itemize}
	\item Serão construídas strings de busca com as palavras-chave e seus sinônimos que estejam relacionados ao objetivo deste processo de revisão sistemática.
	\item As strings serão submetidas aos mecanismos de busca das bases escolhidas.
	\item Após leitura do título, resumo e palavras-chave, serão aplicados critérios de inclusão e exclusão, retirando possíveis duplicatas e extraindo os trabalhos que tenham relevância.
	\item Os artigos serão então lidos na íntegra.
	\item O revisor fará um resumo de cada trabalho selecionado, destacando os métodos utilizados para avaliação e os parâmetros considerados, quando for o caso.
	\\\\\\\\\\\\\\\\\\\\\\\\\\\\\\\\
\end{itemize}

%

\section{Procedimento de execução}
\subsection{Seleção dos Estudos}
\begin{figure}[htbp]
	\centering 
	\includegraphics[width=1.2\columnwidth]{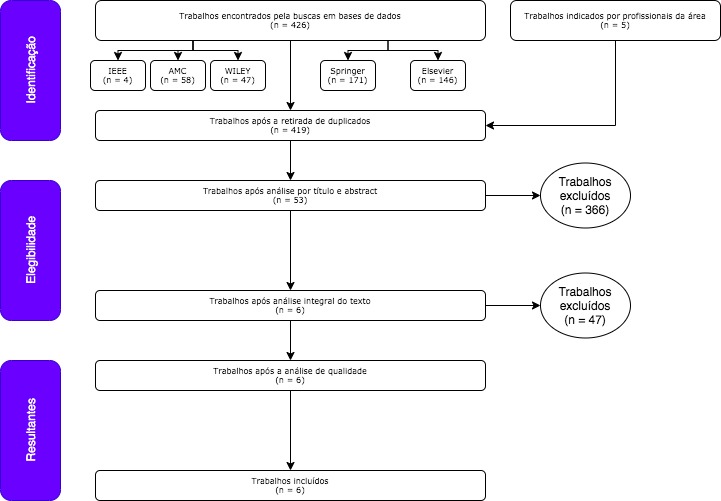} 
	\caption[Ilustração do processo de execução inicial]{Ilustração do processo de execução inicial} 
	\label{fig:gallery_2} 
\end{figure}

\begin{figure}[htbp]
	\centering 
	\includegraphics[width=1\columnwidth]{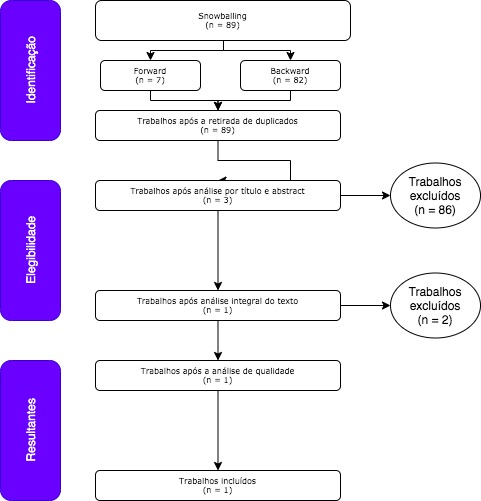} 
	\caption[Ilustração do processo de execução snowball]{Ilustração do processo de execução snowball} 
	\label{fig:gallery_2_snowball} 
\end{figure}

O processo de execução foi realizado conforme ilustrado abaixo(Figura \ref{fig:gallery_2}), sendo utilizado durante a aplicação dos critérios o procedimento metodológico \textit{snowball}(Figura \ref{fig:gallery_2_snowball}). A realização do processo foi por meio de uma planilha disponibilizada \href{https://docs.google.com/spreadsheets/d/1E3YwFvYL24OJOsb9K8XeamA5fo1JgI2gt41AOf87XRk/edit?usp=sharing}{neste link}.
\subsection{Extração de Dados}
O formulário de extração de dados foi feito por meio de tabela, e seu resultado completo está disponibilizado \href{https://docs.google.com/spreadsheets/d/1szYXF195HhL1Ta0gAVFFlL49prUaLL-XwtX1yoz0yNo/edit?usp=sharing}{neste link}. Toda a extração de dados foi realizada unicamente pelo autor, mas confirmada por um julgador externo.
\subsection{Análise da Qualidade}
A análise da qualidade foi realizada segundo o que foi proposto pelo protocolo definido neste trabalho, e baseia-se na estrutura apresentada por \cite{kitchenham2007guidelines}. As perguntas realizadas foram então respondidas por meio de pontuações, e o resultado completo da análise está disponibilizado \href{https://docs.google.com/spreadsheets/d/1FHiYTJeOL8nRCmchVGDFbAkZI666teznxknFeh5u3pw/edit?usp=sharing}{neste link}. Dentre os trabalhos analisados, todos obtiveram níveis de qualidade satisfatório ou alto, mais especificamente metade deles tiveram nota satisfatória e a outra metade notas altas, sendo uma pontuação classificatória, ou seja, nenhum dos trabalhos analisados foi excluído, mesmo por que todos obtiveram um grau positivo de qualidade.

\section{Análise dos Resultados}
Os resultados obtidos pelo trabalho de Revisão Sistemática da Literatura estão dispostos segundo as perguntas de pesquisa definidas pelo protocolo, identificando de maneira gráfica, quais as principais abordagens adotadas pelos estudos identificados.
\subsection{Como a utilização dos protocolos associados a mecanismos de incentivo financeiro tem sido adotados para WMNs?}
\begin{figure}[htb]
	\centering 
	\includegraphics[width=0.8\columnwidth]{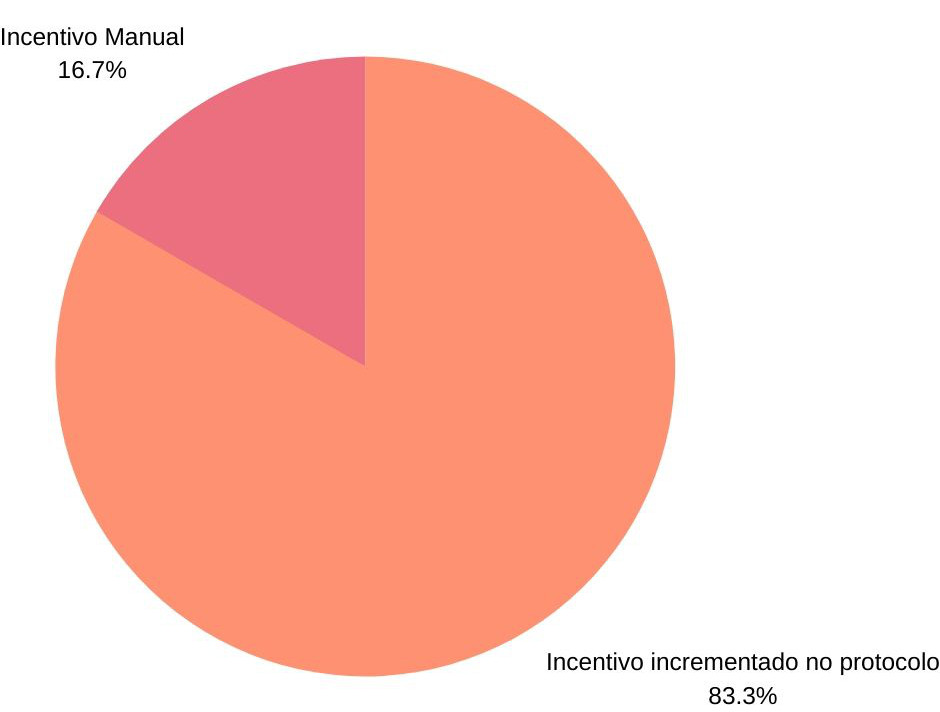} 
	\caption[Tipos de incentivo]{Tipos de incentivo.} 
	\label{fig:tiposIncentivo} 
\end{figure}

A análise feita sobre as formas de utilização dos protocolos demonstrou não apenas os protocolos mais usados, como também uma outra forma de incentivo manual adotada em trabalhos. Como pode ser visto na Figura \ref{fig:tiposIncentivo}, apesar de grande parte das formas de incentivos, aproximadamente 83.3\%, estarem atreladas à modificação de protocolos para inclusão de métricas de preço, ainda existem abordagens que fazem essa precificação de maneira manual, onde o dinheiro é trocado fisicamente entre as pessoas, de acordo com o uso da rede.
 
\begin{figure}[htb]
	\centering 
	\includegraphics[width=0.8\columnwidth]{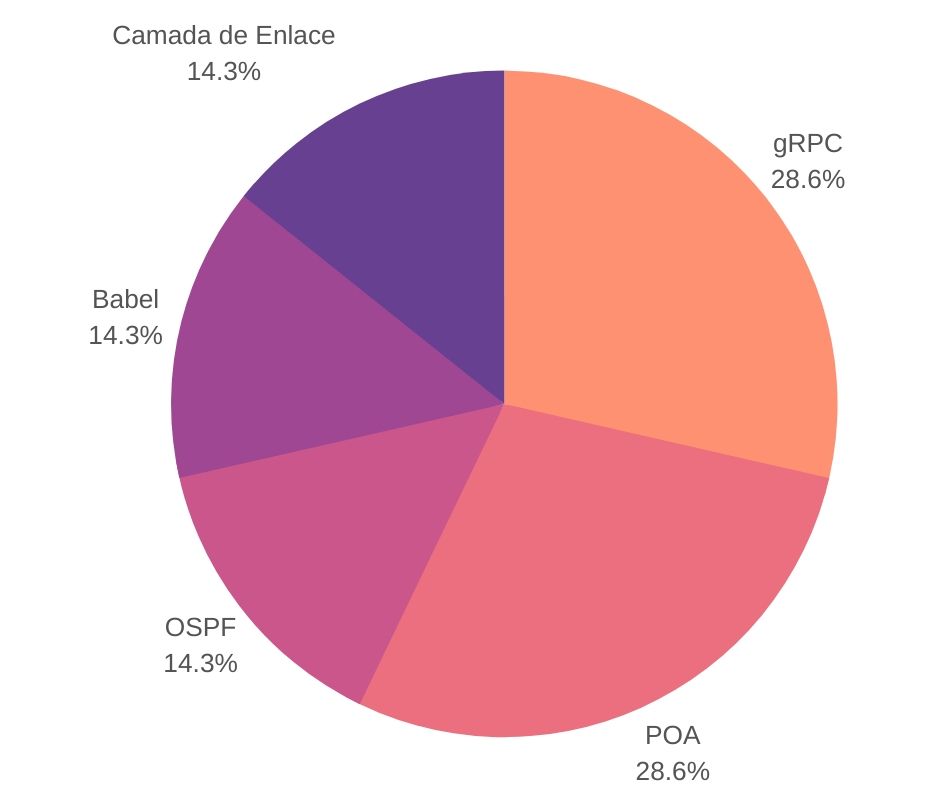} 
	\caption[Protocolos utilizados]{Protocolos utilizados.} 
	\label{fig:protocolos} 
\end{figure}

 Já quando o processo de trocas financeiras é automatizado pelos protocolos, os trabalhos analisados tendem a preferir utilizar os protocolos gRPC e POA, como pode ser visto na Figura \ref{fig:protocolos}. Vale também destacar o protocolo Babel, que apesar de ser utilizado em apenas 14,3\% dos trabalhos, é o protocolo mais recente proposto, e o que apresenta maior eficiência em testes comparativos \cite{7573902}. Outros protocolos também usados como OSPF, já muito conhecido na literatura \cite{moy1998ospf}, além de outros protocolos específicos da camada de enlace, focados nos canais de banda da rede.

\subsection{Quais as técnicas mais utilizadas para implementação de mecanismos de incentivo para WMNs dentro de protocolos de roteamento?}
\begin{figure}[htb]
	\centering 
	\includegraphics[width=0.8\columnwidth]{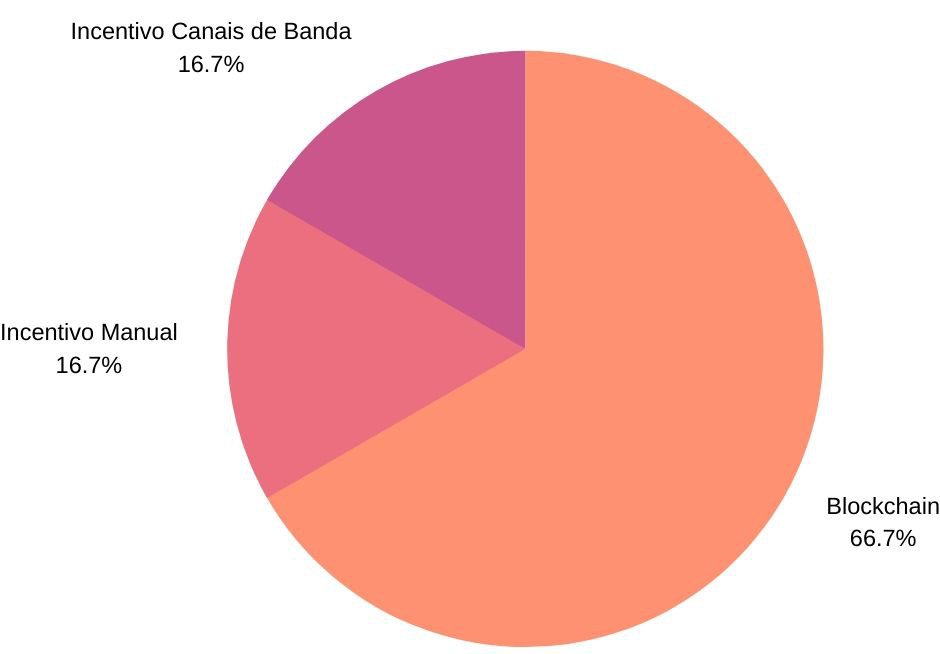} 
	\caption[Técnicas utilizadas pelas propostas]{Técnicas utilizadas pelas propostas.} 
	\label{fig:tecnicas} 
\end{figure}

As técnicas mais utilizadas para inclusão de incentivos estão voltadas para utilização de Blockchains atreladas aos protocolos de rede, onde os protocolos são modificados para incluirem métricas de preço, cobrando ou pagando pelos serviços de encaminhamento de pacotes, e utilizam a tecnologia Blockchain para legitimar e automatizar as trocas financeiras, como pode ser visto na Figura \ref{fig:tecnicas}. Dentro dos trabalhos também foram identificadas outras técnicas, como o incentivo dentro dos canais de banda, precificando cada canal (dentro da camada de enlace), onde quanto maior o congestionamento, mais caro o canal se torna para ser usado; além de incentivos manuais, onde os protocolos apenas analisam o tráfego de rede, mas a cobrança é feita através de contas enviadas aos clientes, de forma manual, como é feita hoje pelas grandes operadoras de cabeamento telefônico.

\subsection{Qual a eficiência dos modelos propostos em testes realizados para o incentivo de WMNs?}
\begin{figure}[htb]
	\centering 
	\includegraphics[width=0.8\columnwidth]{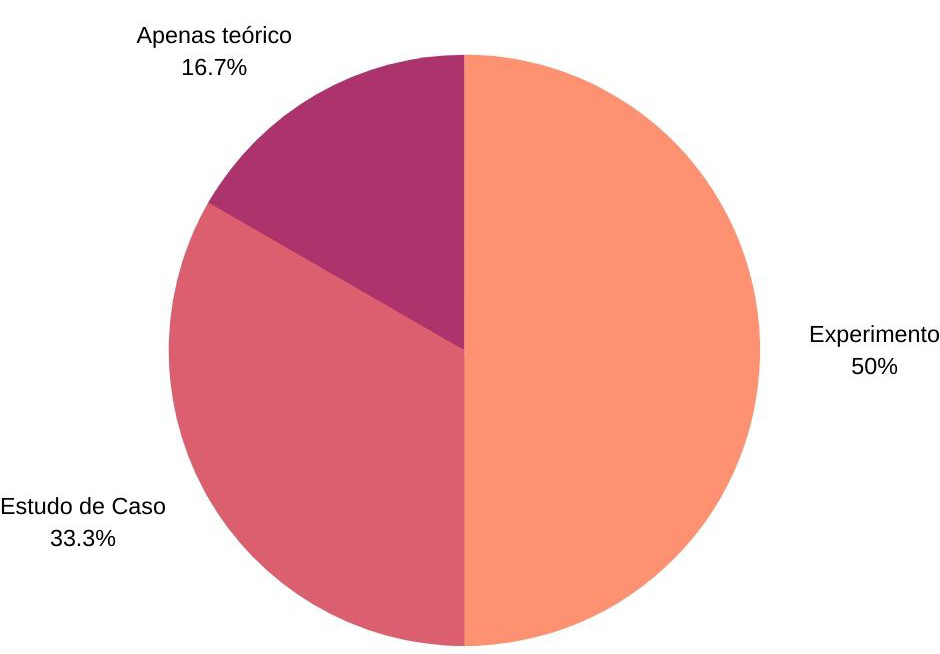} 
	\caption[Tipo de estudo realizado]{Tipo de estudo realizado.} 
	\label{fig:tiposEstudos} 
\end{figure}

A eficiência dos modelos propostos está diretamente ligada a forma como tais modelos foram testados e analisados, mas como podemos ver inicialmente pela Figura \ref{fig:tiposEstudos}, 16,7\% dos trabalhos elencados, ou seja somente 1 trabalho, acabou não tendo sua eficiência evidenciada por meio de testes, enquanto os trabalhos restantes, aproximadamente 83,3\% deles, permitiram que a eficiência dos modelos pudesse ser evidenciada através de estudos de casos reais ou experimentos realizados tanto em laboratório quanto em ambiente de produção.

\begin{figure}[htb]
	\centering 
	\includegraphics[width=0.8\columnwidth]{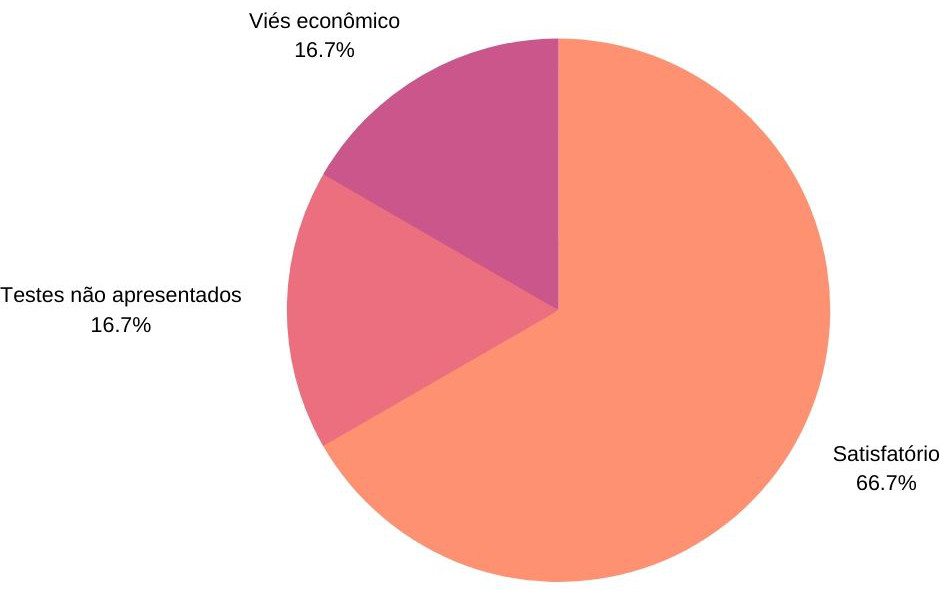} 
	\caption[Eficiência prática das propostas]{Eficiência prática das propostas.} 
	\label{fig:eficiencia} 
\end{figure}

Vale salientar que também foi identificado nos estudos analisados, uma ameaça a validade dos trabalhos que está diretamente relacionada a eficiência de suas propostas, que seria o viés econômico a que os autores estariam expostos, pois ao estarem ligados economicamente à proposta que apresentam, suas análises poderiam deixar de ser imparciais. Diante do exposto, separamos em nossa análise de eficiência, todos os trabalhos que não apresentaram testes práticos ou que os dados pudessem estar enviesados por ligações econômicas, sendo esses uma total de 33,4\% dos trabalhos aproximadamente, como podemos ver pela Figura \ref{fig:eficiencia}. Enquanto que em todos os trabalhos restantes, aproximadamente 66,7\% deles, foi possível identificar um desempenho e eficiência satisfatórios, pois mesmo com alguns problemas de performance vindos da inclusão de incentivo à rede, eles ainda apresentam um ótimo nível de QoS e estabilidade em seus protocolos.
\\\\\\\\\\\\\\\\\\\\\\
\subsection{Quais os benefícios e melhorias podem ser alcançadas para as WMNs com a utilização dos protocolos adaptados com mecanismos de incentivo financeiro?}
\begin{figure}[htb]
	\centering 
	\includegraphics[width=0.75\columnwidth]{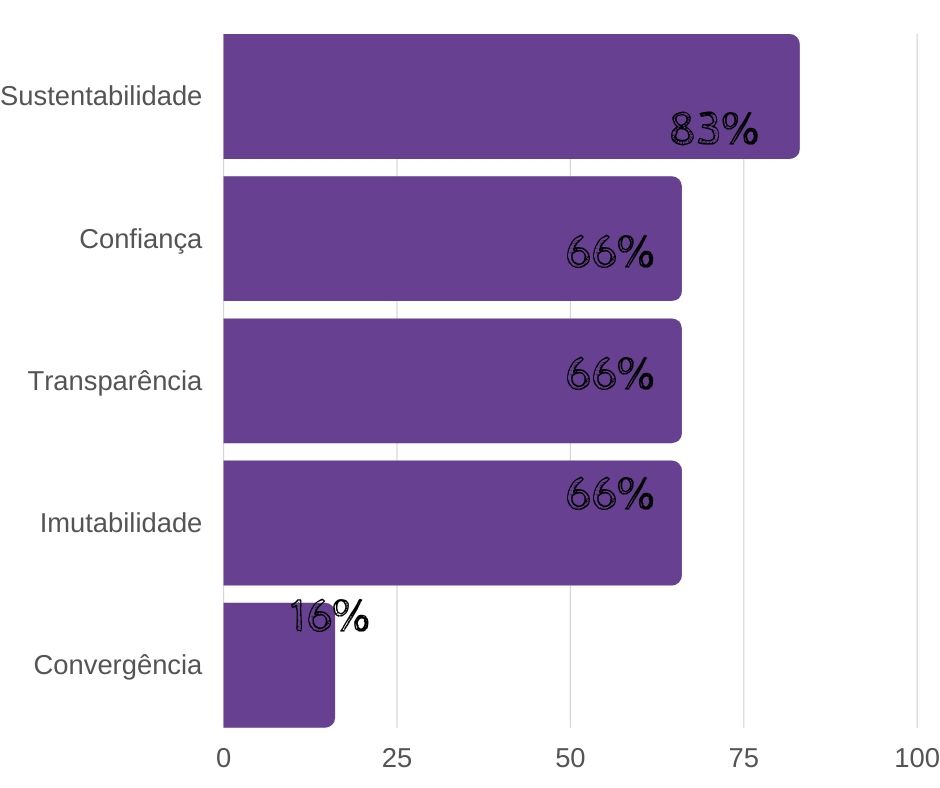} 
	\caption[Benefícios e melhorias identificados]{Benefícios e melhorias identificados.} 
	\label{fig:beneficios} 
\end{figure}

Dentro dos benefícios e melhorias elencados pelos trabalhos, destaca-se a sustentabilidade da rede, citado em 83\% dos estudos (Figura \ref{fig:beneficios}), já que os mecanismos de incentivo tem como principal objetivo, garantir um retorno viável para quem mantém a rede e age para contribuir com o crescimento da mesma. Outros benefícios são também elencados como confiança entre os participantes, transparência das transações e imutabilidade dos dados que as guardam, garantidas pela utilização de tecnologias como a Blockchain. Também foram citados benefícios na melhora do tempo de convergência, mas tal característica ocorre apenas em propostas que lidam com protocolos a nível de enlace, pois permite atribuir métricas que incentivem a escolha dos melhores links de comunicação, de maneira mais ágil e assertiva, já em protocolos a nível de rede, como é abordado na maior parte dos estudos analisados, a convergência não costuma ter melhoras, pois as métricas adicionam processamento que costuma consumir banda de rede, levando a uma melhora de convergência de poucos trabalhos, em aproximadamente 16\% deles apenas.

\subsection{Quais os custos e riscos associados a utilização dos protocolos adaptados com mecanismos de incentivo financeiro para as WMNs?}
\begin{figure}[htb]
	\centering 
	\includegraphics[width=0.75\columnwidth]{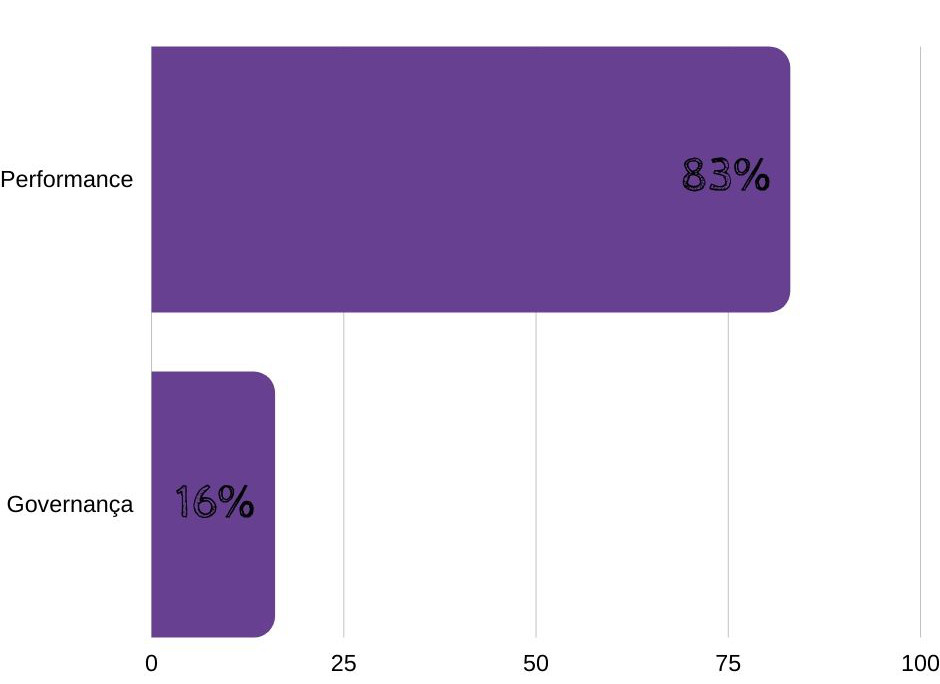} 
	\caption[Custos e riscos identificados]{Custos e riscos identificados.} 
	\label{fig:riscos} 
\end{figure}

Os trabalhos analisados apresentam como problemas, diversos desafios que costumam estar atrelados a dois aspectos, a performance da rede e as regras de governança da mesma (Figura \ref{fig:riscos}). As redes WMNs, como costumam ser executadas em dispositivos limitados, tanto em processamento como em energia, e a inclusão de novas métricas a serem analisadas causa uma maior demanda de processamento, que pode afetar a performance a nível de tráfego e banda de redes. Como esse impacto na performance costuma ser aceitável segundo testes, e por aumentar as funcionalidades principais de incentivo e estabilidade da rede, as propostas tronam-se viáveis. A execução dos protocolos, por ser sempre atrelada ao meio físico, costuma sofrer certas interferência dos mesmos, e dependendo de aspectos como localização dos nós e forma de pagamento dos participantes, podem existir problemas caso a rede não tenha um nível de governança mínimo e eficaz, capaz de manter a ordem e qualidade da mesma.
\section{Conclusão}
Através da análise realizada dos estudos apresentados, é possível identificar a exploração de redes de malha sem fio como sendo uma possibilidade real e implementável, capaz de se tornar globalmente aplicável quando em conjunto com tecnologias como blockchain juntamente com políticas de recompensa por meio de criptomoedas(ou outras forma que sejam de interesse dos participantes da rede), além de garantir mecanismos básicos de segurança que permitam maior confiabilidade entre os integrantes da rede, segurança nas transações de recompensa entre os participantes da rede e diminuição dos custos de armazenamento e produção de blocos para a borda da rede. Mas ainda existem muitos desafios a serem alcançados, tanto em questões de segurança e privacidade da rede, quanto em problemas atrelados à performance e governança da rede, que ainda podem ser mais explorados dentro de possíveis trabalhos futuros.
\\\\\\\\\\\\\\\\

\renewcommand{\refname}{\spacedlowsmallcaps{Referências}} 

\bibliographystyle{unsrt}

\bibliography{sample.bib} 


\end{document}

%% file: structure.tex
\usepackage[
nochapters, 
beramono, 
eulermath,
pdfspacing, 
dottedtoc 
]{classicthesis} 

\usepackage{arsclassica} 

\usepackage[T1]{fontenc} 

\usepackage[utf8]{inputenc} 

\usepackage{graphicx} 
\graphicspath{{Figures/}} 

\usepackage{enumitem} 

\usepackage{lipsum} 

\usepackage{subfig} 

\usepackage{amsmath,amssymb,amsthm} 

\usepackage{varioref} 

\theoremstyle{definition} 

\theoremstyle{plain} 

\theoremstyle{remark} 

\hypersetup{
colorlinks=true, breaklinks=true, bookmarks=true,bookmarksnumbered,
urlcolor=webbrown, linkcolor=RoyalBlue, citecolor=webgreen, 
pdftitle={}, 
pdfauthor={\textcopyright}, 
pdfsubject={}, 
pdfkeywords={}, 
pdfcreator={pdfLaTeX}, 
pdfproducer={LaTeX with hyperref and ClassicThesis} 
}